# PASIR RIS SECONDARY SCHOOL

**DEVELOPMENT OF A JAVA PACKAGE**
**FOR MATRIX PROGRAMMING**
**PTB/PRSS/2002/001**

**TEACHER FACILITATOR:** **MRS LIM NGEE PENG**

**EXTERNAL FACILITATOR:** **MR LING HAN TONG MAURICE**

**STUDENTS:** **LIM YUE CHUANG SHAWN (S8726921C)**
**CHOI JI HEE (F2492071L)**
**TEO BOON KOK HENRY (S8727457H)**

# CONTENTS

# Acknowledgement

We would like to thank Mrs Lim Ngee Peng, for introducing such an interesting and enriching project.

We would like to thank Mr Maurice Ling, for his guidance and teaching us Java programming

We would like to thank Mr Teo Pat, for his assistance in mathematics.

Lastly, we would like to thank all our friends, who had helped us in one way or another.

# ABSTRACT

We had assembled a Java package, known as MatrixPak, of four classes for the purpose of numerical matrix computation. The classes are matrix, matrix_operations, StrToMatrix, and MatrixToStr; all of which are inherited from java.lang.Object class. Class matrix defines a matrix as a two-dimensional array of float types, and contains the following mathematical methods: transpose, adjoint, determinant, inverse, minor and cofactor. Class matrix_operations contains the following mathematical methods: matrix addition, matrix subtraction, matrix multiplication, and matrix exponential. Class StrToMatrix contains methods necessary to parse a string representation (for example, [[2 3 4]-[5 6 7]]) of a matrix into a matrix definition, whereas class MatrixToStr does the reverse.

# Introduction

We used Java for our program is because it is an easier language to use and is more widely used. Because that Java has a library to use from, we are able to do simple basic algebraic calculations. Java is widely used over the internet, so people from all over the world is able to gain access to our program and thus we would be able to aid more people. Java uses applet, which allows users to do matrix calculations over the internet. So this is why we used Java as the computer language to write our program.

This program is all about matrix. We wrote it to help people through out the world to solve matrices problems that are simply too hard and long to do. We did this program on matrix are for various reasons, firstly, for someone to do matrix calculations in large sums, example a 10 X 10 matrix, it would be very tedious, takes up a long time and could easily make a mistake. Second, matrices are more frequently used by engineers, commuters and a salesman, the matrices that they have to solve would be a lot bigger than a 10 X 10 matrix, so it would be nearly impossible for them to it, and if they really could, it would take a very long time. So we did this program as a remedy to these problems. Our program could also be used for students, teachers and many others, it could simply be used to solve simple matrices for students, or it could be used to do complicated matrices.

Our program has is able to execute many different operations, they are addition, subtraction, multiplication, transpose, inverse, cofactor, determinant, adjoint and power. So, our program gives our users much help as operations like determinant and power are very tedious and long to do.

# LITERATURE REVIEW

## Historical computer algebra systems

Much of the attempts of symbolic and algebraic manipulation of mathematical equations using computer pre-dates the 1970s. A stream of emphasis on computer algebra systems then were towards symbolic solutions rather than numerical solutions as seen today. This emphasizes uses computers as automation, automating routine mathematics.

Many questions arise from using computers as mathematical symbolic manipulators. One of the tougher questions can be worded as follows: Given the large variety and complexity of equations, how can it be tackled effectively? No doubt, even till today there's no one system capable of handling all equations. Nevertheless, there are works in this area. By the work of Kamke [KAM61] in the late 1950s, he generalized the number of types differential equations into 367 forms, and demonstrates the treatment of each type. This formed the basic of certain systems in the 1970s, such as, that of Schmidt's [SCH76] EULE.

Other programs developed during that period of time or prior to EULE include, Moses' SOLDIER [MOS67], Slagle's SAINTS [SLA63], and Wang's WANDERER [WAN71].

Much of the programmes mentioned earlier uses heuristics as symbolic solutions are greatly heuristics such process. In EULE [SCH76], differential equations are treated by the methods described by Murphy [MUR60], with the knowledge presented by Kamke [KAM61]. It first attempts to simplify the given equations before searching for a method of solution.

Slapler's SAINT (symbolic automatic INTegrator) [SLA63] started as part of the doctoral these of the author in Massachusetts Institute of Technology (MIT) [SLA61]. Its main procedure is by constructing a tree and pruning from it at the same time until a solution is reached, which had been previously described [NEW57]. At each stage, cost is calculated for OR trees and costly branches are pruned.

## Symbolic Input

In all computer mathematical systems, one of the essential and implicit problems to encounter is the mean of symbolic input. How can equation be entered into a computer via mainly the keyboard? How will these inputs be entered and understood? The problem arises as mathematical notations are two-dimensional scripts and computers can only read one-dimensional scripts. Until later years where GUT and mouse become common, the keyboard is the sole input device. Thus, how to fit a two-dimensional scripting into a single dimension, without losing its implicit meanings, becomes a thing to solve.

The first electronic computer was created to reduce time in calculation. It was until the late 1970s that they had other agenda. Hence its sole purpose then was in mathematics. Computer programmes, even today, are essentially nothing more than lists of formulas. In the 1950s, work was dome in creating a language known as FORTRAN (FORmula TRANslator) for such a purpose. In fact, FORTRAN was the first programming language that was standardized, known as FORTRAN66 [FOR66], since it was 1966. FORTRAN66 underwent two major revisions in the last century, resulting in FORTRAN77 [FOR77] and FORTRAN90 [FOR90].

However, there were many symbolic input methods and schemes that did not fall into the mainstream as FORTRAN. They were developed as more specialized ways by different groups to attack specific areas of mathematics. At the same time more general schemes were also developed. Early methods were mainly text-based, such as Ossenbruggen's [OSS71], Barton's [BAR70], AUTOMAST [BAL66] and SCRATCHPAD [JEN74]. From 1990, symbolic inputs were more graphical, following closely to the original notations and WYSIWYG philosophy. Examples included Zhao's [ZHA96], Lamagna's [LAM92], SUI [DOL90], MathScribe [SMI86], and internet-friendly notations, such as, OpenMath [PRI00].

Despite being textual systems, early systems have great degree of variance in the input methods. Ossenbruggen [OSS71] uses a rather structure less model as compared to Barton [BAR70], which requires a programmatic structure in based on total Autocode [BAR67]. In this aspect it is similar to FORTRAN, in the spectrum, AUTOMAST [BAL66] requires a semi-structure in its input, which is of much lesser stringency than that on Barton's.

SCRATCHPAD [JEN74] is implemented as an interactive semi-programming environment. Being so, there are defined structures and syntax. However, it can also be seen as a communication schemes there are essential no processing behind the interface. SCRATCHPAD is this, an elaborated user interface.

The box language presented by Zhao's [AHA96] were based on former research by the group [ZHA94] [SAK96] but added the features of symbol omission for brevity. Symbol omissions may be determinable (usually by the order of operations) or indetermination (resulting in ambiguous notations). Undeterminable omissions are often due to context and personal habits but are clear to the writer. Box language presents this by requiring all [arts of the equation to be casted in boxes. Thus, computable clarity is maintained, together with personal habits.

Lamagna's [LAM92] user interface was to be used in Newton, as an additional package to Maple. The interface checks for mathematical correctness at each stage of input and amendment.

MathScribe [SMI86] is based on a series of prior works [LEL85] [FOS84] [MAR71] of graphical interfaces. MathScribe is similar to the box language described by Zhao's [ZHA96]. However, MathScribe portrays less stringency in the box language is highly structured to provide mathematical clarity whereas MathScribe relies on the natural mathematical representation and only inherits the box-like structure from it. Hence, in MathScribe, there is no explicit definition of the box.

Mathematica is a commercial suite that uses a relative of box language, known as Bra-ket Notation [HAR00]. Bra-Ket Notation is implemented as "function-like" methods. With resemblance to function calls in many programming languages, each box has a defined name and parameters. These functions or boxes can either play formatting roles, like superscript functions or boxes can either like Laplace Transform. Bra-Ket Notation adapts a linear scheme of input, which will mean that readability suffers.

OpenMath [PRI00] is one of the two major moves of moving natural mathematical notation into the Internet. The other being MathML (Mathematical Markup Language). It mimics the implementation of HTML [RAG99]. Due to the nature and the length of the document produced in either OpenMath or MathML, native readability suffers greatly and often, error detection is impossible without an able browser.

## Matrix Operations[1]

For the purpose of this project, the following matrix operations are considered: addition, subtraction, multiplication, power, transposition, minor, cofactor, adjoint, inverse and determinant.

Given two matrices, $A = [a_{i,j}]$ and $B = [b_{i,j}]$, where i is the row position and j is the column position of the elements in the matrix. For this section, the above definition and notation maintains. Under circumstances where the number of rows and columns of A and B are respectively equal, matrix addition is defined as $[a_{i,j} + b_{i,j}]$. The resulting matrix will have the same row and column size as A or B.

Given two matrices, $A = [a_{i,j}]$ and $B = [b_{i,j}]$, under circumstances where the number of rows and columns of A and B are respectively equal, matrix subtraction is defined as $[a_{i,j} - b_{i,j}]$. The resulting matrix will have the same row and column size as A or B.

[1] Information for this section is gathered from the following sources, [AYR74], [BOU02], [HAE99].

Given two matrices, A = [$a_{i,j}$] and B = [$b_{i,j}$], under circumstances where the number of rows of A and the number of columns of B are equal, matrix multiplication is defined as [Σ( $a_{i,j}$ x $b_{i,j}$ )] (row multiplied by column). Another method is to transpose matrix A, and then do a column multiplied by column operation. The product of two matrices will have the number of rows of A and the number of columns as B. Power or exponential operation on matrices is similar to multiplication of scalar values.

Given a matrix, A = [$a_{i,j}$], the transpose of A is defined as, $A^T$ = [$a_{j,i}$]. In simple terms, the rows become the columns and the columns become the rows.

The minor of a matrix is a reduced size matrix. A minor is denoted as |$M_{i,j}$|, and it reduces the size of the original matrix (assumed to be M x N) to (M-1) x (N-1) matrix by deleting row i and column j of the original matrix.

Cofactor is also known as signed minor. As the name implies, it is an extension from a minor. Hence a cofactor is defined as $(-1)^{i+j}$ $a_{i,j}$ |$M_{i,j}$|.

An adjoint of a matrix is defined as the transposition of cofactor matrix.

Inverse of a matrix is defined as the scalar division of the adjoint matrix by the determinant of the matrix.

Given a 2 x 2 matrix where row 1 is [a,b] and row 2 is [c,d], the determinant is defined as (ad – bc). In the case of a higher order matrix, it is then reduced to a 2 x 2 matrix via recursive cofactoring operation.

## Java Programming Language Features[2]

Java is a programming language platform from Sun Microsystems. It is a simpler langauge as compared to C or C++. The language is object oriented. By using Java, your development cycle is much faster because Java technology is interpreted. The compile-link-load-test-crash-debug cycle is obsolete. Now you just compile and run. Your applications are portable across multiple platforms. Write your application once and you never need to port them - they will run without modification on multiple operating systems and hardware architectures. Your applications are robust because the Java runtime environment managers memory for you. Your interactive graphical applications have high performance because multiple concurrent threads of activity in your application are supported by the multithreading built into the Java programming language and runtime platform. Your applications are adoptable to changing environments because you can dynamically download code from anywhere on the network. Your end users can trust that your applications are secure, even through that downloading code from all over the internet; the Java runtime environment has built-in protection against viruses and malicious codes. So the Java programming platform provides a portable, interpreted, high-performance, simple, object-oriented programming language and supporting runtime environment.

[2] Information for this section is gathered from the following source, [GOS96]

Java is simple as it can be programmed without extensive programming training while being attached to current software practices. The fundamental concepts of Java technology are grasped quickly; programmers can be productive from the very beginning.

Java is object oriented, the needs of distributed, client-server based systems coincide with the encapsulated, message-passing paradigm of object-based software. To function within increasingly complex network-based environments, programming systems must adopt object-oriented concepts. Java technology provides a clean and efficient object-based development platform.

Java is a similar language as it looks like C++. Java is easier as it removes the complications of C++. Having the Java programming language retaining many of the object-oriented features and the "look and feel" of C++ means that programming can migrate easily to the Java platform and be producing quickly.

Java is Robust as it was designed for creating high reliability. It provides extensive compile-time checking, followed by a second level of runtime checking.

Java is secure as it was designed to operate in distributed environments, which means that security is of paramount importance. High security features designed into the language and runtime system, Java technology lets you complete applications written in the Java programming language are safe from intension by unauthorized code file systems.

Java is designed for network usage as the java complier produce bytecodes on architecture neutral intermediate format designed to transport code efficiently to multiple hardware and software platforms.

Java is portable as it was the language platform of Java technology known as Java Virtual Machine.

Java has high performance as it adopts a scheme by which the interpreter can run at full speed without needing to check the runtime environment. It has the automatic garbage collector to ensure high memory.

Java is an interpreted language. The Java interpreter can execute Java bytecode directly as any machine to which the interpreter and runtime system has been ported.

Java technology's multithreading capability provides the means to built applications with many concurrent threads of activity. Multithreading thus results in a high degree of interactivity for the end user.

Java is dynamic as the language and runtime system are dynamic in their linking stages.

# FORMAL SPECIFICATION

## Problem Formalization

Given (M x N) matrices where M = N or M ≠ N, we seek a library in Java programming language, which holds matrix operations for use for pre-undergraduate calculations and experimentations.

## Solution Formalization

```
matrix = [mArray : N  x  N  ;
          numofrows : N  ;
          numofcols : N  ]
```

We define a matrix as a two-dimensional array of float.

```
Constructor01 = [D  matrix ;
                mArray? :N  x  N  ;
                numofrows? : N  ;
                numofcols?:N  |
                numofrows? > 1 ;
                numofcols? > 1 ;
                mArray a mArray? ;
                numofrows a numofrows? ;
                numofcols a  numofcols ]
```

```
Constructor02 = [D  matrix ;
                numofrows? : N ;
                numofcols?: N  |
                numofrows? > 1 ;
                numofcols? > 1 ;
                numofrows a  numofrows? ;
                numofcols a  numofcols? ]
```

```
Fillmatrix = [D matrix ;
             mArray? N  x  N  |
             numofrows > 1 ;
             numofcols > 1 ;
             mArray a mArray? ]
```

isSquare = [ return! : **N** |
(numofrows = numofcols) **f** (return! = 1) ;
(numofrows <> numofcols) **f** (return! = 0)]

isIdentity = [ return! : **N** |
isSquare() **!** = 1 **f** (return! = 2);
**W** (**N**i = 1) **!** = 1 ^ **W** (**N**i,i = 1) **!** = 1 **f** (return! = 1);
**W** (**N**i = 1) = 1 ^ **W** (**N**i,i = 1) = 1 **f** (return! = 0)]

addition = [sum! : matrix ;
addeum? : matrix ;
adder? : matrix |
numofrows **.** addeum? = numofrows **.** adder?;
numofcols **.** addeum? = numofcols **.** adder?;
sum! (**N**i , **N**j) = addeum? (**N**i , **N**j) + adder? (**N**i , **N**j)]

subtraction = [diff! : matrix ;
subtor? : matrix ;
subdeum?: matrix |
numofrows **.** subtor? = numofrows **.** subdeum? ;
numofcols **.** subtor? = numofcols **.** subdeum? ;
diff! (**N**i , **N**j) = subdeum? (**N**i , **N**j) - subtor? (**N**i , **N**j)]

multiplication = [product! : matrix ;
multor? : matrix ;
multiplent? : matrix |
numofrows **.** multiplent? = numofcols **.** . multor? ;
transpose (multiplent?) ;
product! (**N**i , **N**j) = **e** multiplent? (**N**i , **N**j) **x** (multor?(**N**i , **N**j)]

power = [product! : matrix ;
multiplent? : matrix ;
inc? > **N** | ;
product! = multiplent?$^{inc?}$ ]

transpose = [transposition! : matrix ;
m? : matrix |
transposition! (**N**i , **N**j) = m? (**N**j , **N**i)]

minor = [ minor! : matrix;
m? : matrix;
row? : **N**;
col? : **N** |
row? ^ col? > 0;
row? <= Max($m?_{row}$);
col? <= Max($m?_{col}$);
((row? ^ col? = 1) ➔ (minor!($\mathbf{N}_{i=0}^{i=\text{Max(m?row)-1}}$, $\mathbf{N}_{j=0}^{j=\text{Max(m?col)-1}}$)
= m?($\mathbf{N}_{i=1}^{i=\text{Max(m?row)}}$, $\mathbf{N}_{j=1}^{j=\text{Max(m?col)}}$))) **v**
((row? = Max($m?_{row}$) ^ col? = 1) ➔
(minor!($\mathbf{N}_{i=0}^{i=\text{Max(m?row)-1}}$, $\mathbf{N}_{j=0}^{j=\text{Max(m?col)-1}}$)
= m?($\mathbf{N}_{i=0}^{i=\text{Max(m?row)-1}}$, $\mathbf{N}_{j=1}^{j=\text{Max(m?col)}}$))) **v**
((row? = 1 ^ col? = Max($m?_{col}$)) ➔
(minor!($\mathbf{N}_{i=0}^{i=\text{Max(m?row)-1}}$, $\mathbf{N}_{j=0}^{j=\text{Max(m?col)-1}}$)
= m?($\mathbf{N}_{i=1}^{i=\text{Max(m?row)}}$, $\mathbf{N}_{j=0}^{j=\text{Max(m?col)-1}}$))) **v**
((row? = Max($m?_{row}$) ^ col? = Max($m?_{cow}$)) ➔
(minor!($\mathbf{N}_{i=0}^{i=\text{Max(m?row)-1}}$, $\mathbf{N}_{j=0}^{j=\text{Max(m?col)-1}}$)
= m?($\mathbf{N}_{i=0}^{i=\text{Max(m?row)-1}}$, $\mathbf{N}_{j=0}^{j=\text{Max(m?col)-1}}$))) **v**
(minor!($\mathbf{N}_{i=0}^{i=\text{Max(m?row)-1}}$, $\mathbf{N}_{j=0}^{j=\text{Max(m?col)-1}}$)
= m?($\mathbf{N}_{i=0}^{i=\text{row?-1}}$, $\mathbf{N}_{j=0}^{j=\text{col?-1}}$) ^
m?($\mathbf{N}_{i=0}^{i=\text{row?)-1}}$, $\mathbf{N}_{j=\text{col?+1}}^{j=\text{Max(m?col)}}$) ^
m?($\mathbf{N}_{i=\text{row?+1}}^{i=\text{Max(m?row)}}$, $\mathbf{N}_{j=0}^{j=\text{col?-1}}$) ^
m?($\mathbf{N}_{i=\text{row?+1}}^{i=\text{Max(m?row)}}$, $\mathbf{N}_{j=\text{col?+1}}^{j=\text{Max(m?col)}}$))]

cofactor = [ cofactor! : matrix;
m? : matrix |
Max($m?_{row}$) = Max($m?_{col}$);
cofactor!($\mathbf{N}_i$, $\mathbf{N}_j$) **a**
determinant(minor(m? >> m?, i >> row?, j >> col?), i >> row?) *
$(-1)^{i+j}$]

adjoint = [ adjoint! : matrix;
m? : matrix |
Max($m?_{row}$) = Max($m?_{col}$);
adjoint! **a** transpose(cofactor(m?))]

inverse = [ inverse! : matrix;
m? : matrix |
Max($m?_{row}$) = Max($m?_{col}$);
inverse! **a** adjoint(m?) / determinant(m?, Max($m?_{row}$))]

determinant = [ determinant! : **N**;
col? : **N**;
m? : matrix |
$Max(m?_{row}) = Max(m?_{col})$;
(col? = 1) ➔ determinant! = m?($\mathbf{N}_0$, $\mathbf{N}_0$);
(col? = 2) ➔ determinant! = m?($\mathbf{N}_0$, $\mathbf{N}_0$) * m?($\mathbf{N}_1$, $\mathbf{N}_1$) –
m?($\mathbf{N}_0$, $\mathbf{N}_1$) * m?($\mathbf{N}_1$, $\mathbf{N}_0$);
(col? > 2) ➔ determinant! =
determinant(minor(m? >> m?, 1 >> row?, col? >> $col?)_{col?=1}^{col?=col?}$
>> m?, col? – 1)]

*quod erat demonstrandum*

# SYSTEM DESIGN

## Matrix definition

Computationally, we define a numerical matrix as a two-dimensional array of float elements, with the number of rows and the number of columns of the array stored as discrete variables.

Mechanically, a matrix representation is entered into the library as String of characters. Each elements in the row is separated by a space, for example, “3 4 5”. Each row is definitely enclosed by “[“ and “]”, for example “[3 4 5]. Each row is linked consecutively by “-“, for example, “[3 4 5]-[6 7 8]”. Finally, the entire representation is flanked be “[“ and “]”, for example, “[[3 4 5]-[6 7 8]]”.

## To change from a string to matrix

Matrix is typed as a string in this from, example -> [[2 3 4]-[2 3 4]-[2 3 4]].
We clear everything before and after the first and last brackets.
The first and last open and close brackets are the matrix brackets. The other open and close brackets [2 3 4] are the rows. The ‘-‘ is used to put the columns.

Now, to change it to array form, firstly, we have to get rid of the first and last brackets. Then we have [2 3 4]-[2 3 4]-[2 3 4]. Then we break them into an array of elements [2 3 4], [2 3 4] and [2 3 4]. We count the number of ‘-‘ and add one to be the number of rows in the matrix. Next we get rid of all the other brackets. But it is still in string form, so we break the strings out the white spaces in between. We change it into string buffer, then to string and finally to float. After all these are dome, we will get our matrix.

## Transpose

Transposing a matrix is simply to change a matrix’s row into column and column into row. By doing so in java, we take it that each number in the matrix like (1,2), has a coordinate point. Example ‘I’ in the matrix has a coordinate point of (0,0) and 2 has a coordinate point of (1,0) and so on. So to flip the row to column and column to rows, we flip the coordinate point. Example to transpose (1,2) into (1,3), the coordinates of '2', (0,1) would change to (1,0) and so its position in the matrix is flipped. In Java, we have 'i' for rows and 'j' for columns, so it is (i,j). After we flipped the matrix, we would get (j,i).

Now in Java form, we have two matrix, m and temp.m. By using Array, I for row and j for column, we put, temp.mArray[j][i] = mArray[i][j]. Then we return temporary matrix.

Example:

$$\begin{bmatrix} 2 & 5 \\ 3 & 4 \\ 4 & 3 \\ 5 & 2 \end{bmatrix}^T = \begin{bmatrix} 2 & 3 & 4 & 5 \\ 5 & 4 & 3 & 2 \end{bmatrix}$$

$$\begin{bmatrix} 2 & 7 & 5 & 1 & 5 \\ 2 & 8 & 4 & 1 & 2 \\ 4 & 9 & 3 & 4 & 6 \\ 6 & 5 & 2 & 4 & 4 \end{bmatrix}^T = \begin{bmatrix} 2 & 2 & 4 & 6 \\ 7 & 8 & 9 & 5 \\ 5 & 4 & 3 & 2 \\ 1 & 1 & 4 & 4 \\ 5 & 2 & 6 & 4 \end{bmatrix}$$

## Addition

For addition, two matrix p and q are taken in and returns object m.

When there's a difference in row size and column size, it throws Arithmetic Exception to show that there's an error.

When there are no errors, matrix m is created as a result (new object).

We use two if statements to check row and column.

When row size is not equal, statement showing error is thrown.

When column size is not equal, statement of showing error is thrown.

When p.rowsize is equivalent to q.rowsize, for loop is used to find the coordinate points of the integer using i and j.

Using matrix addition law, i of matrix p is added to i of matrix j.

Since m is created as a new object, m is the end result of the addition and returns to the main.

Example:

$$\begin{bmatrix} 2 & 5 & 2 & 4 \\ 3 & 4 & 2 & 6 \\ 4 & 3 & 1 & 5 \\ 5 & 2 & 1 & 3 \end{bmatrix} + \begin{bmatrix} 5 & 2 & 2 & 4 \\ 4 & 3 & 6 & 2 \\ 3 & 4 & 1 & 5 \\ 2 & 5 & 3 & 3 \end{bmatrix} = \begin{bmatrix} 7 & 7 & 4 & 8 \\ 7 & 7 & 8 & 8 \\ 7 & 7 & 2 & 10 \\ 7 & 7 & 4 & 6 \end{bmatrix}$$

## Subtraction

For subtraction, two matrices p and q is taken in and returns object m.

Matrix m is a new object that has the same row and column size with p.

Two if statements are used to check for row size and column size.

If p.rowsize is not equivalent to q.rowsize, Arithmetic Exception is thrown in to declare that there’s a difference in row size.

If p.colsize is not equivalent to q.colsize, Arithmetic Exception is thrown in to declare that there’s a difference in column size.

When there are same row and column size, for loop is used to find the coordinate points using i and j, by using i++, which increases by 1 for every loop. Until I reaches the row size, for loop is used to find the coordinate of j and increases by 1 in every loop.

M, the new matrix object stores the value of end result of subtraction of p and q which is done according to matrix subtraction law.

M is returned to the main.

Exanple:

$$\begin{bmatrix} 2 & 5 & 2 & 4 \\ 3 & 4 & 2 & 6 \\ 4 & 3 & 1 & 5 \\ 5 & 2 & 1 & 3 \end{bmatrix} - \begin{bmatrix} 2 & 2 & 4 & 2 \\ 4 & 3 & 6 & 2 \\ 4 & 3 & 5 & 1 \\ 2 & 5 & 3 & 1 \end{bmatrix} = \begin{bmatrix} 0 & 3 & -2 & 2 \\ -1 & 1 & -4 & 4 \\ 0 & 0 & -4 & 4 \\ 3 & -3 & -2 & 2 \end{bmatrix}$$

## Power

In this function, it takes in matrix p, integer n, where n is the value of power.

Since it is in public, it can be called by any other classes when necessary.

If statement is used to check when n is equivalent to 1, it returns to original matrix p as in all cases when the power equals to 1, the value remains the same.

Else is used to check when n is not equivalent to 1. If statement is used to check when n=2, by calling multiplication class, p is multiplied by p (itself), which means p^2. The product matrix is then returned to the main.

Another else statement is used to check for n is neither equivalent to 1 nor 2.

When n is neither 1 nor 2, by using multiplication class, integer n is reduced by 1 and p is multiplied by itself according to the result of n-1 times before multiplying p again.

For example, when n=3,

(p,power(p,(3)-1);

where p; (p.2) which means p;(p(p)) where p;(p,2) which means p;(p(p)). Hence the value becomes p*p*p which is equivalent to p^3.

Example:

$$\begin{bmatrix} 3 & 7 & 4 \\ 5 & 8 & 1 \\ 6 & 3 & 2 \end{bmatrix}^3 = \begin{bmatrix} 3 & 7 & 4 \\ 5 & 8 & 1 \\ 6 & 3 & 2 \end{bmatrix} x \begin{bmatrix} 3 & 7 & 4 \\ 5 & 8 & 1 \\ 6 & 3 & 2 \end{bmatrix} x \begin{bmatrix} 3 & 7 & 4 \\ 5 & 8 & 1 \\ 6 & 3 & 2 \end{bmatrix}$$

$$= \begin{bmatrix} 68 & 95 & 27 \\ 61 & 102 & 30 \\ 45 & 72 & 31 \end{bmatrix} x \begin{bmatrix} 3 & 7 & 4 \\ 5 & 8 & 1 \\ 6 & 3 & 2 \end{bmatrix}$$

$$= \begin{bmatrix} 841 & 1317 & 421 \\ 873 & 1333 & 406 \\ 345 & 984 & 314 \end{bmatrix}$$

## Extraction of minor from p matrix by an element

There are fine cases in getting the minor of the matrix, they are, (0.0) ; (row, 0) ; (0, col) ; (row, col) ; (row-m, col-n).

By using a 10 by 10 matrix as an example, for the first case, to get the minor, a 9 by 9 matrix, we would have to eliminate the first row and first column. So now, we have a 9 by 9 matrix, but the coordinate points of the first number would be (1,1). Because that Java array starts with (0,0), so we would need to change the matrix to start from 0,0. To do that, all we have to do is to -1 for row and column, then we would have a correct 9 by 9 matrix.

Now, for the second case, (row,0) which means the last row and first column, we have to eliminate them. By using a 10 by 10 matrix as an example, we want to get the minor of it, a 9 by 9 matrix. After we eliminate the last row and first column, the coordinates now have to be moved to the left which means that the column all must be -1. so after doing that we would get a correct 9 by 9 and 10 by 10 matrix.

Now, for the third case, (0,col), we would have to get rid of the first row and last column. By using a 10 by 10 matrix as an example, we want to get the minor, a 9 by 9 matrix. We after eliminating them, we would need to move the matrix up so that, it would start from (0,0). To do that, we would -1 for the row of all the coordinate points. Then we would get a correct 9 by 9 matrix.

For the forth case r (row, col), we have to eliminate the last row and last column. Using a 10 by 10 matrix 0,0 as an example, we want to get the minor of it, a 9 by 9 matrix. After eliminating the last row and last column, we would get a correct 9 by 9 matrix straight away because the coordinates starts from (0,0) already.

For case 5, we need to eliminate (n,m). n is the row somewhere between the first and last and m is also the column somewhere between the first and the last.

By using a 10 by 10 matrix as an example, we want to get a 9 by 9 matrix which is a minor of it. To do that, we eliminate n row and m column, after that, 10 by 10. We have to paste the four parts, labeled 1, 2, 3 and 4 on the diagram together. The first part 1, we do not need to do anything since it is already in its correct position. So for the second part, 2, we have to move it to the left in the diagram, so to do that we -1 for the column. For the forth part, we have to -1 for its row and columns. Then we would get a correct 9 by 9 matrix.

Example:

$$\begin{bmatrix} 2 & 7 & 3 & 2 & 3 & 8 \\ 4 & 9 & 3 & 4 & 9 & 4 \\ 5 & 1 & 1 & 8 & 1 & 1 \\ 6 & 5 & 6 & 5 & 2 & 7 \\ 8 & 5 & 1 & 3 & 8 & 7 \end{bmatrix}$$

$$M1,1 = \begin{bmatrix} 9 & 3 & 4 & 9 & 4 \\ 1 & 1 & 8 & 1 & 1 \\ 5 & 6 & 5 & 2 & 7 \\ 5 & 1 & 3 & 8 & 7 \end{bmatrix}$$

$$M5,1 = \begin{bmatrix} 7 & 3 & 2 & 3 & 8 \\ 9 & 3 & 4 & 9 & 4 \\ 1 & 1 & 8 & 1 & 1 \\ 5 & 6 & 5 & 2 & 7 \end{bmatrix}$$

$$M3,3 = \begin{bmatrix} 2 & 7 & 2 & 3 & 8 \\ 4 & 9 & 4 & 9 & 4 \\ 6 & 5 & 5 & 2 & 7 \\ 8 & 5 & 3 & 8 & 7 \end{bmatrix}$$

## Multiplication

For matrix multiplication, the number of columns of the first matrix and the number of rows of the second matrix must be equal, and the product has a size of the number of rows of the first matrix and the number of columns of the second matrix. Regularly, matrix multiplication requires the summation of the product of row elements and column elements. However, for computational simplicity, it is changed to the summation of product of column elements of the first matrix and the column elements of the second matrix. In order to do so, transposition of the first matrix is required. Hence, summation of the product of the column elements of the transposed first matrix and the column elements of the second matrix yields the element of the first row and first column of the product (resultant) matrix. In another words, upon transposition of the first matrix, the column number of the transposed matrix will determine the row position of the product, whereas the column number of the second matrix will determine the column position.

Example:

$$\begin{bmatrix} 2 & 7 \\ 5 & 8 \\ 6 & 9 \end{bmatrix} \begin{bmatrix} 3 & 4 & 7 & 6 \\ 2 & 8 & 7 & 1 \end{bmatrix} = \begin{bmatrix} 20 & 64 & 63 & 19 \\ 31 & 84 & 91 & 38 \\ 36 & 96 & 109 & 45 \end{bmatrix}$$

## Determinant

First, we need to find that if it is a 2 by 2 matrix, if it is, then we would have (2x6 – 4x5). If the matrix is not a 2 by 2 matrix, we would create n number of arrays; each array would have a minor of n-1. Then for the n-1 matrix, we would repeat what was done until we get n = 2. When we want to get the determinant, we need to add the minors up, to do that, each minor, we need to add them and subtract them alternately. Then we would get the determinant.

Example:

$$\left| \begin{bmatrix} 2 & 5 \\ 4 & 6 \end{bmatrix} \right| = (2x6 - 4x5) = -8$$

$$\begin{vmatrix}\begin{bmatrix}7 & 2 & 9 & 2\\ 8 & 1 & 5 & 2\\ 9 & 4 & 9 & 3\\ 5 & 6 & 1 & 7\end{bmatrix}\end{vmatrix} = (-1)^{1+1}(7)\begin{bmatrix}1 & 5 & 2\\ 4 & 9 & 3\\ 6 & 1 & 7\end{bmatrix} + (-1)^{1+2}(2)\begin{bmatrix}8 & 5 & 2\\ 9 & 9 & 3\\ 5 & 1 & 7\end{bmatrix} + (-1)^{1+3}(9)\begin{bmatrix}8 & 1 & 2\\ 9 & 4 & 3\\ 5 & 6 & 7\end{bmatrix}$$

$$+ (-1)^{1+4}(2)\begin{bmatrix}8 & 1 & 5\\ 9 & 4 & 9\\ 5 & 6 & 1\end{bmatrix}$$

$$= (-1)^{1+1}(7)\left((-1)^{1+1}(1)\begin{bmatrix}9 & 3\\ 1 & 7\end{bmatrix} + (-1)^{1+2}(5)\begin{bmatrix}4 & 3\\ 6 & 7\end{bmatrix} + (-1)^{1+3}(2)\begin{bmatrix}4 & 9\\ 6 & 1\end{bmatrix}\right)$$

$$+ (-1)^{1+2}(2)\left((-1)^{1+1}(8)\begin{bmatrix}9 & 3\\ 1 & 7\end{bmatrix} + (-1)^{1+2}(5)\begin{bmatrix}9 & 3\\ 5 & 7\end{bmatrix} + (-1)^{1+3}(2)\begin{bmatrix}9 & 9\\ 5 & 1\end{bmatrix}\right)$$

$$+ (-1)^{1+3}(9)\left((-1)^{1+1}(8)\begin{bmatrix}4 & 3\\ 6 & 7\end{bmatrix} + (-1)^{1+2}(1)\begin{bmatrix}9 & 3\\ 5 & 7\end{bmatrix} + (-1)^{1+3}(2)\begin{bmatrix}9 & 4\\ 5 & 6\end{bmatrix}\right)$$

$$+ (-1)^{1+4}(2)\left((-1)^{1+1}(8)\begin{bmatrix}4 & 9\\ 6 & 1\end{bmatrix} + (-1)^{1+2}(1)\begin{bmatrix}9 & 9\\ 5 & 1\end{bmatrix} + (-1)^{1+3}(5)\begin{bmatrix}9 & 4\\ 5 & 6\end{bmatrix}\right)$$

$$= 322$$

## Cofactor

To get the cofactor of a matrix, firstly, we let the matrix seem to have minor matrices within itself. Then in each individual minor matrix, we find the determinant. This step could also be called signed minor. We are actually getting the signed minor of the minor matrix. Then after we got it for each individual matrix, we put them together and get back a 4 by 4 matrix which is the cofactor matrix.

Example:

$$\begin{bmatrix} 7 & 2 & 9 & 2 \\ 8 & 1 & 5 & 2 \\ 9 & 4 & 9 & 3 \\ 5 & 6 & 1 & 7 \end{bmatrix} = \begin{bmatrix} \alpha_{1,1} & \alpha_{1,2} & \alpha_{1,3} & \alpha_{1,4} \\ \alpha_{2,1} & \alpha_{2,2} & \alpha_{2,3} & \alpha_{2,4} \\ \alpha_{3,1} & \alpha_{3,2} & \alpha_{3,3} & \alpha_{3,4} \\ \alpha_{4,1} & \alpha_{4,2} & \alpha_{4,3} & \alpha_{4,4} \end{bmatrix}$$

$$= \begin{bmatrix} (-1)^{1+1}(7)\,|M_{1,1}| & (-1)^{1+2}(2)\,|M_{1,2}| & (-1)^{1+3}(9)\,|M_{1,3}| & (-1)^{1+4}(2)\,|M_{1,4}| \\ (-1)^{2+1}(8)\,|M_{2,1}| & (-1)^{2+2}(1)\,|M_{2,2}| & (-1)^{2+3}(5)\,|M_{2,3}| & (-1)^{2+4}(2)\,|M_{2,4}| \\ (-1)^{3+1}(9)\,|M_{3,1}| & (-1)^{3+2}(4)\,|M_{3,2}| & (-1)^{3+3}(9)\,|M_{3,3}| & (-1)^{3+4}(3)\,|M_{3,4}| \\ (-1)^{4+1}(5)\,|M_{4,1}| & (-1)^{4+2}(6)\,|M_{4,2}| & (-1)^{4+3}(6)\,|M_{4,3}| & (-1)^{4+4}(7)\,|M_{4,4}| \end{bmatrix}$$

$$= \begin{bmatrix} (-1)^{1+1}(7)\left|\begin{bmatrix} 1 & 5 & 2 \\ 4 & 9 & 3 \\ 6 & 1 & 7 \end{bmatrix}\right| (-1)^{1+2}(2)\left|\begin{bmatrix} 8 & 5 & 2 \\ 9 & 9 & 3 \\ 5 & 1 & 7 \end{bmatrix}\right| (-1)^{1+3}(9)\left|\begin{bmatrix} 8 & 1 & 2 \\ 9 & 4 & 3 \\ 5 & 6 & 7 \end{bmatrix}\right| (-1)^{1+4}(2)\left|\begin{bmatrix} 8 & 1 & 5 \\ 9 & 4 & 9 \\ 5 & 6 & 1 \end{bmatrix}\right| \\ (-1)^{2+1}(8)\left|\begin{bmatrix} 2 & 9 & 2 \\ 4 & 9 & 3 \\ 6 & 1 & 7 \end{bmatrix}\right| (-1)^{2+2}(1)\left|\begin{bmatrix} 7 & 9 & 2 \\ 9 & 9 & 3 \\ 5 & 1 & 7 \end{bmatrix}\right| (-1)^{2+3}(5)\left|\begin{bmatrix} 7 & 2 & 2 \\ 9 & 4 & 3 \\ 5 & 6 & 7 \end{bmatrix}\right| (-1)^{2+4}(2)\left|\begin{bmatrix} 8 & 1 & 5 \\ 9 & 4 & 9 \\ 5 & 6 & 1 \end{bmatrix}\right| \\ (-1)^{3+1}(9)\left|\begin{bmatrix} 2 & 9 & 2 \\ 1 & 5 & 2 \\ 6 & 1 & 7 \end{bmatrix}\right| (-1)^{3+2}(4)\left|\begin{bmatrix} 7 & 9 & 2 \\ 8 & 5 & 2 \\ 5 & 1 & 7 \end{bmatrix}\right| (-1)^{3+3}(9)\left|\begin{bmatrix} 7 & 2 & 2 \\ 8 & 1 & 2 \\ 5 & 6 & 7 \end{bmatrix}\right| (-1)^{3+4}(3)\left|\begin{bmatrix} 7 & 2 & 9 \\ 8 & 1 & 5 \\ 5 & 6 & 1 \end{bmatrix}\right| \\ (-1)^{4+1}(5)\left|\begin{bmatrix} 2 & 9 & 2 \\ 1 & 5 & 2 \\ 4 & 9 & 3 \end{bmatrix}\right| (-1)^{4+2}(6)\left|\begin{bmatrix} 7 & 9 & 2 \\ 8 & 5 & 2 \\ 9 & 9 & 3 \end{bmatrix}\right| (-1)^{4+3}(1)\left|\begin{bmatrix} 7 & 2 & 2 \\ 8 & 1 & 2 \\ 9 & 4 & 3 \end{bmatrix}\right| (-1)^{4+4}(7)\left|\begin{bmatrix} 7 & 2 & 9 \\ 8 & 1 & 5 \\ 9 & 4 & 9 \end{bmatrix}\right| \end{bmatrix}$$

$$(-1)^{1+1}(7)\left((-1)^{1+1}(1)\begin{bmatrix}9&3\\1&7\end{bmatrix}+(-1)^{1+2}(5)\begin{bmatrix}4&3\\6&7\end{bmatrix}+(-1)^{1+3}(2)\begin{bmatrix}4&9\\6&1\end{bmatrix}\right)$$

$$(-1)^{1+2}(2)\left((-1)^{1+1}(8)\begin{bmatrix}9&3\\1&7\end{bmatrix}+(-1)^{1+2}(5)\begin{bmatrix}9&3\\5&7\end{bmatrix}+(-1)^{1+3}(2)\begin{bmatrix}9&9\\5&1\end{bmatrix}\right)$$

$$(-1)^{1+3}(9)\left((-1)^{1+1}(8)\begin{bmatrix}4&3\\6&7\end{bmatrix}+(-1)^{1+2}(1)\begin{bmatrix}9&3\\5&7\end{bmatrix}+(-1)^{1+3}(2)\begin{bmatrix}9&4\\5&6\end{bmatrix}\right)$$

$$(-1)^{1+4}(2)\left((-1)^{1+1}(8)\begin{bmatrix}4&9\\6&1\end{bmatrix}+(-1)^{1+2}(1)\begin{bmatrix}9&9\\5&1\end{bmatrix}+(-1)^{1+3}(5)\begin{bmatrix}9&4\\5&6\end{bmatrix}\right)$$

$$(-1)^{2+1}(8)\left((-1)^{1+1}(2)\begin{bmatrix}9&3\\1&7\end{bmatrix}+(-1)^{1+2}(9)\begin{bmatrix}4&3\\6&7\end{bmatrix}+(-1)^{1+3}(2)\begin{bmatrix}4&9\\6&1\end{bmatrix}\right)$$

$$(-1)^{2+2}(1)\left((-1)^{1+1}(7)\begin{bmatrix}9&3\\1&7\end{bmatrix}+(-1)^{1+2}(9)\begin{bmatrix}9&3\\5&7\end{bmatrix}+(-1)^{1+3}(2)\begin{bmatrix}9&9\\5&1\end{bmatrix}\right)$$

$$(-1)^{2+3}(5)\left((-1)^{1+1}(7)\begin{bmatrix}4&3\\6&7\end{bmatrix}+(-1)^{1+2}(2)\begin{bmatrix}9&3\\5&7\end{bmatrix}+(-1)^{1+3}(2)\begin{bmatrix}9&4\\5&6\end{bmatrix}\right)$$

$$(-1)^{2+4}(2)\left((-1)^{1+1}(7)\begin{bmatrix}4&9\\6&1\end{bmatrix}+(-1)^{1+2}(2)\begin{bmatrix}9&9\\5&1\end{bmatrix}+(-1)^{1+3}(9)\begin{bmatrix}9&4\\5&6\end{bmatrix}\right)$$

$$(-1)^{3+1}(9)\left((-1)^{1+1}(2)\begin{bmatrix}5&2\\1&7\end{bmatrix}+(-1)^{1+2}(9)\begin{bmatrix}1&2\\6&7\end{bmatrix}+(-1)^{1+3}(2)\begin{bmatrix}1&5\\6&1\end{bmatrix}\right)$$

$$(-1)^{3+2}(4)\left((-1)^{1+1}(7)\begin{bmatrix}5&2\\1&7\end{bmatrix}+(-1)^{1+2}(9)\begin{bmatrix}8&2\\5&7\end{bmatrix}+(-1)^{1+3}(2)\begin{bmatrix}8&5\\5&1\end{bmatrix}\right)$$

$$(-1)^{3+3}(9)\left((-1)^{1+1}(7)\begin{bmatrix}1&2\\6&7\end{bmatrix}+(-1)^{1+2}(2)\begin{bmatrix}8&2\\5&7\end{bmatrix}+(-1)^{1+3}(2)\begin{bmatrix}8&1\\5&6\end{bmatrix}\right)$$

$$(-1)^{3+4}(3)\left((-1)^{1+1}(7)\begin{bmatrix}1&5\\6&1\end{bmatrix}+(-1)^{1+2}(2)\begin{bmatrix}8&5\\5&1\end{bmatrix}+(-1)^{1+3}(9)\begin{bmatrix}8&1\\5&6\end{bmatrix}\right)$$

$$(-1)^{4+1}(5)\left((-1)^{1+1}(2)\begin{bmatrix}5&2\\9&3\end{bmatrix}+(-1)^{1+2}(9)\begin{bmatrix}1&2\\4&3\end{bmatrix}+(-1)^{1+3}(2)\begin{bmatrix}1&2\\4&3\end{bmatrix}\right)$$

$$(-1)^{4+2}(6)\left((-1)^{1+1}(7)\begin{bmatrix}5&2\\9&3\end{bmatrix}+(-1)^{1+2}(9)\begin{bmatrix}8&2\\9&3\end{bmatrix}+(-1)^{1+3}(7)\begin{bmatrix}1&2\\4&3\end{bmatrix}\right)$$

$$(-1)^{4+3}(1)\left((-1)^{1+1}(7)\begin{bmatrix}1&2\\4&3\end{bmatrix}+(-1)^{1+2}(2)\begin{bmatrix}8&2\\9&3\end{bmatrix}+(-1)^{1+3}(2)\begin{bmatrix}8&1\\9&4\end{bmatrix}\right)$$

$$(-1)^{4+4}(7)\left((-1)^{1+1}(7)\begin{bmatrix}1&5\\4&9\end{bmatrix}+(-1)^{1+2}(2)\begin{bmatrix}8&5\\9&9\end{bmatrix}+(-1)^{1+3}(9)\begin{bmatrix}8&1\\9&4\end{bmatrix}\right)$$

$$=\begin{pmatrix}-530&-336&900&388\\-70&-84&-2210&56\\477&868&-369&-654\\-145&-576&1&532\end{pmatrix}$$

## Adjoint

To get the adjoint, we simplify transpose the cofactor matrix and we would have our adjoint matrix.

Example:

$$\begin{bmatrix} 7 & 2 & 9 & 2 \\ 8 & 1 & 5 & 2 \\ 9 & 4 & 9 & 3 \\ 5 & 6 & 1 & 7 \end{bmatrix} = \begin{bmatrix} -630 & -336 & 900 & 388 \\ -70 & -84 & -210 & 56 \\ 477 & 868 & -369 & -654 \\ -145 & -576 & 1 & 532 \end{bmatrix}^T = \begin{bmatrix} -630 & -70 & 477 & -145 \\ -336 & -84 & 868 & -576 \\ 900 & -210 & -369 & 1 \\ 388 & 56 & -654 & 532 \end{bmatrix}$$

## Inverse

To get the inverse of a matrix, we multiply each element in the adjoint matrix by the determinant.

Example:

$$\begin{bmatrix} 7 & 2 & 9 & 2 \\ 8 & 1 & 5 & 2 \\ 9 & 4 & 9 & 3 \\ 5 & 6 & 1 & 7 \end{bmatrix}^{-1} = \frac{1}{322}\begin{bmatrix} -630 & -70 & 477 & -145 \\ -366 & -84 & 868 & -576 \\ 900 & -210 & -369 & 1 \\ 388 & 56 & -654 & 532 \end{bmatrix} = \begin{bmatrix} -1.957 & -0.217 & 1.481 & -0.45 \\ -1.043 & -0.261 & 2.696 & -1.789 \\ 2.795 & -0.652 & -1.146 & 0.003 \\ 1.205 & 0.174 & -2.031 & 1.652 \end{bmatrix}$$

# FUTURE WORK

This project has achieved the following:

- Defining a computational entity for numerical matrix operations
- Creating conversion routines from String to matrix entity
- Creating conversion routines from matrix to String entity
- Assembling of the following numerical matrix methods:
    1. Transposition
    2. Determination
    3. Adjoin of matrix
    4. Inversion
    5. Cofactoring
    6. Extraction of minor
    7. Matrix addition
    8. Matrix subtraction
    9. Matrix (regular) multiplication
    10. Matrix exponential

However, the library developed this work is largely untested and is not standalone. Hence much future work on it is possible. Nevertheless, this work forms the basis of a number of possible uses, such as, development into a full calculator or developing into an educational aid.

Future work includes:

- Developing an applet interface
- Enhance the compactness and universality of the library
- Add support for algebraic manipulations
- Increasing the repertoire of functions, such as, eigenvalues and eigenvectors

# BIBLIOGRAPHY

[AYR74] Ayres, Frank, Jr. 1974. Schaum's Outline of Theory and Problems of Matrices, SI (Metric) Edition. McGraw-Hill Book Company.

[BAR66] Ball, W.E and Berns, R.L. 1966. AUTOMAST: Automatic Mathematical Analysis and Symbolic Translation. Communications f the ACM 9,8. Pg 626-633.

[BAR67] Barron, D, Brown, H, Hartley, D and Swinnerton-Dyer, HPF. 1967. Titan Autocode Programming Manual. University Mathematical Laboratory, Cambridge.

[BOU02] Bourke, Paul. 2002. Inverse of a Square Matrix. http://astronomy.swin.edu.au/~pbourke/analysis/inverse/

[CAR02] Carlisle, D, Ion, P, Miner, R, Diaz, A, Hunter, R, Smith, B, Soiffer, N, Sutor, R and Watt, S. 2002. Mathematical Markup Language (MathML) Version2.0 (2nd Edition). W3C working Draft 19 December 2002.

[DOL90] Doleh, Y and Wang, PS. 1990. SUI: A System Independent User Interface for an Integrated Scientific Computing Environment. Proceedings of he International Symposium in Symbolic and Algebraic Computation. Pg. 88-143.

[FOR66] Fortran 66 Standard. 1966. ANSI C3.9-1966.

[FOR77] Fortran 77 Standard. 1977. ISO 1539-1980

[FOS84] Foster.G. 1984. DREAMS: Display Representation for Algebraic Manipulation Systems. UCB/CSB 84/193. University of California, Berkeley.

[FOR90] Fortran 90 Standard. 1990. ISO 1539: 1991 and ANSI X3.198-1992

[GOS96] Gosling, James, and McGilton, Henry. 1996. The Java Language Environment: A White Paper.

[HAE99] Haeussler, Ernest F., Jr. and Paul, Richard S. 1999. Matrix Algebra. In Introductory Mathematical Analysis for Business, Economics, and the Life and Social Sciences, 9th Edition. Pg 209 – 284.

[HAR00] Harris. J. 2000. Advanced Notations in Mathematica. Proceedings of the 2000 International Symposium on Symbolic and Algebraic Computation. Pg 153-160.

[JEN74] Jenks. RD. 1974. The SCRATCHPAD Language. Proceedings of the ACM SIGPLAN Symposium on Very High level languages. Pg. 101-111.

[KAM61] Kambe, E.1961. Differentialglichungen- Loningsmethoden und Losungen, I.Gewohnliche Differentialgleichungen. Alcad Verl. Ges. Geest & Portig k.G.Leipzig 7th edition

[LAM92] Lamagna. EA. Hayden, MB and Johnson , CW. 1992. The Design if a user interface to a computer A;gebra System for Introductory Calcuus. Proceedings of the 1992 International Symposium on Symbolic and Algebraic Computation . Pg 358-368.

[LEL85] Leler. W and Soiffer. N. 1985. An Interactive Graphical Interface for Reduce. ACM SIGSAM Bulletin 19,3. Pg 17-23

[MAR71] Martin. A. 1971. Computer Input/ Output of Mathematical Expressions. Proceedings of the 2nd Symposium in Symbolic and Algebraic Manipulation. Pg. 78-89.

[MOS67] Moses, J. 1967. Symbolic Integration. Report MAC TR-47. Project MAC, MIT, Cambridge.

[MUR60] Murphy, GM. 1960. Ordinary Differential Equations. D.van Nostrand Comp.Inc. New Jersey.

[NEW57] Newell, A, Shaw, JC. and Sumon, HA. 1957. Emperical Explorations of the Logic Theory Machine. Proceedings of the Western Joint Computing Conference. Pg 218-239

[OSS71] Ossenbruggen, PJ, Wong, AKC and Au, T.1971. Symbolic and Algebraic Manipulation for Solving and Differential Equations. Proceedings of the 1971 Annual conference. Pg. 717-735.

[PRI00] Prieto, H, Dalmas, S and Papegay, Y.2000. Mathematica as an OpenMath Application. ACM SIGSAM 34, 2. Pg 22-26

[RAG99] Raggett, D. 1999. HTML 4.01 Specification. WSC Recommendation 24 December 1999.

[SAK96] Sakurai, T, Zhao, Y, Sugiura, H and Torii, T. 1996. A User Interface for Natural Mathematical Notations. Transactions of the Japanese Society for Industrial and Applied Mathematics 6,1. Pg 147-157.

[SLA61] Slagler, JR. 1961. A Heuristic Program that solves Symbolic Integration Problems in Freshman Calculus: Symbolic Automatic Integrator (SAINT). Report 5 G-0001. MIT, Lincoln Laboratory, Cambridge.

[SLA63] Slagle, JR. 1963. A Heuristic Program that solves Symbolic Integration Problems in Freshman Calculus, Journal of the ACM, 10. Pg 507-520

[SMI86] Smith. CJ and Soiffer, NM. 1986. MathScribe: AUuser interface for Computer Algebra Systems. Proceedings of the 5th ACM Symposium on Symbolic and Algebraic Computation. Pg. 7-12

[WAN71] Wang, PS. 1971. Evaluation of Definite Integrates by Symbolic Manipulation. Report MAC TR-92. Project MAC, MIT, Cambridge.

[ZHA94] Zhao, Y, Sugiura, H, Torti, T and Saleurai, T. 1994. A Knowledge-Based Method for Mathematical Notations Understanding. Transactions of Information Processing Society of Japan 35, 11. Pg. 2366-2381.

[ZHA96] Zhao, YJ, Sakurai, T, Sugiura, H and Torii, T. 1996. A Methodology of Parsing Mathematical Notations for Mathematical Computation. Pg. 292-300.

# APPENDIX A:
# PACKAGE SOURCE CODES

```
/**
 * matrix.java
 *
 * This file contain class matrix.
 * It forms the foundational ADT (the matrix) and all its operations.
 * Class matrix holds all the operations on single matrix.
 *
 * Created on January 30, 2003, 4:47 AM
 */

/**
 *
 * @author  Maurice@Server
 */
import java.lang.ArrayIndexOutOfBoundsException;
import java.lang.IndexOutOfBoundsException;
import java.lang.NegativeArraySizeException;
import java.lang.ArithmeticException;
import java.lang.IllegalArgumentException;
import java.lang.Exception;
import java.lang.Math;
import java.lang.Number;
import java.lang.Float;
import java.lang.Double;

public class matrix extends java.lang.Object {

    /**
    A class to contain a matrix and all operations dealing with itself. A matrix is any 2-
dimensional array of float numbers.
    */
    int numofrows = 0;
    int numofcols = 0;
    float[][] mArray;

    // THIS SECTION CONTAINS CONSTRUCTORS AND ASSIGNMENT
METHODS.
    public matrix(int row, int col) throws NegativeArraySizeException
```

```
    /**
    description: constructor method. produces an undeclared matrix array of given rows
and columns
    */
    {
        if (row < 0)
        {
            throw new NegativeArraySizeException("matrix(int row, int col) error. row < 0");
        };
        if (col < 0)
        {
            throw new NegativeArraySizeException("matrix(int row, int col) error. col < 0");
        };
        mArray = new float[row][col];
        numofrows = row;
        numofcols = col;
    }

    public matrix(float[][] m, int row, int col) throws ArrayIndexOutOfBoundsException,
NegativeArraySizeException
    /**
    description: constructor method. produces a matrix array of given rows and columns
and declares its elements as given in the parameters.
    */
     {
         if (row < 0)
        {
            throw new NegativeArraySizeException("matrix(float[][] m, int row, int col)
error. row < 0");
        };
        if (col < 0)
        {
            throw new NegativeArraySizeException("matrix(float[][] m, int row, int col)
error. col < 0");
        };
        if (m.length != row || m[1].length != col)
        {
            // print error message if row or column not equals to m's row and column
            System.out.println("Matrix constructor error: allocation size different. public
matrix(float[][] m, int row, int col)");
            throw new ArrayIndexOutOfBoundsException("matrix(float[][] m, int row, int
col) size mismatch");
        }
        else
        {
            mArray = new float[row][col];
```

```
            mArray = m;
            numofrows = row;
            numofcols = col;
        };
    }

    public void fillmatrix(float[][] m) throws ArrayIndexOutOfBoundsException
    /**
    description: assign m to the matrix
    */
    {
        if (m.length != numofrows || m[1].length != numofcols)
        {
            // print error message if row or column not equals to m's row and column
            System.out.println("Matrix constructor error: allocation size different. public void
fillmatrix(float[][] m)");
            throw new ArrayIndexOutOfBoundsException("fillmatrix(float[][] m) assign
error");
        }
        else
        {
            mArray = m;
        }
    }
    // END OF CONSTRUCTOR AND ASSIGNMENT METHODS.

    // THIS SECTION CONTAINS INFORMATION METHODS.
    public int rowsize()
    /**
    description: returns the number of rows of this matrix
    */
    {
        return numofrows;
    }

    public int colsize()
    /**
    description: returns the number of columns of this matrix
    */
    {
        return numofcols;
    }

    public float[][] showMatrix()
    /**
     * description: exports the contents of a matrix as an array
```

```
 */
{
   return mArray;
};
// END OF INFORMATION METHODS.

// THIS SECTION CONTAINS CHECKING / INFORMATION METHODS.
public int isSquare()
/**
description: checking method, to see if this matrix is a square matrix
return: returns 1 if this is square
        returns 0 if this is not square
*/
{
   if (numofrows == numofcols)
      return 1;
   else
      return 0;
}

public int isIdentity()
/**
 *description: checking method, to see of this matrix is identity
 *return: returns 1 if this is identity
 *        returns 0 if this is not identity
 *        returns 2 if this is not square (shouldn't even be checking)
 */
{
   int[] presence = new int[numofrows];
   int tester = 1;
   // tester = 1 means identity
   int count = 0;
   if (isSquare() == 0)
   {
      return 2;
   }
   else
   {
      /* until this part, mArray is a square matrix.
       *only if the entire row has only one "1" will presence[i] = 1
       *if it is a null row, presence[i] will be 0
       *if the row has more than one non-zero, presence[i] > 1
       */
      for (int i = 0; i < numofrows; i = i + 1)
      {
         presence[i] = 0;
```

```
            for (int j = 0; j < numofrows; j = j + 1)
            {
                if (mArray[i][j] == 0)
                {
                    presence[i] = presence[i] + 0;
                }
                else
                {
                    presence[i] = presence[i] + 1;
                };
            };
        };
        while (count < numofrows)
        {
            if (presence[count] == 1)
            {
                count = count + 1;
            }
            else
            {
                tester = 0;
                count = count + 1;
            };
        };
        if (tester == 1)
        {
            return 1;
        }
        else
        {
            return 0;
        }
    }
}
// END OF CHECKING / INFORMATION METHODS.

// THIS SECTION CONTAINS OPERATION METHODS.
public matrix transpose()
/**
description: method for matrix multiplication
parameters: matrix
return: 1 2-dimensional matrix containing the transposition
algorithm:
*/
{
    matrix temp = new matrix(numofcols, numofrows);
```

```
        for (int i = 0; i < numofrows; i = i + 1)
        {
            for (int j = 0; j < numofcols; j = j + 1)
            {
                temp.mArray[j][i] = mArray[i][j];
            };
        };
        return temp;
    }

    public matrix extract_minor(int row, int col) throws IndexOutOfBoundsException
    /**
    description: extraction of 1 minor from p matrix by an element
    parameter: element by row and column
    return M-1 x N-1 matrix
    algorithm: create (numofrows - 1, numofcols - 1) matrix
     5 possible cases: (0,0); (row,0); (0,col); (row,col); (row - m, col - n).
     *handles each of the possible cases as follows:
     *for (0,0) do from (1 to row, 1 to col)
     *for (row,0) do (1 to row - 1, 0 to col)
     *for (0, col) do (0 to row, 1 to col - 1)
     *for (row, col) do (0 to row - 1, 0 to col - 1)
     *for (m, n) do (0 to m - 1, 0 to n - 1)
     *            (0 to m - 1, n + 1 to col)
     *            (m + 1 to row, 0 to n - 1)
     *            (m + 1 to row, n + 1 to col)
    */
    {
        matrix minor = new matrix(numofrows - 1, numofcols - 1);
        if (row > numofrows)
        {
            throw new IndexOutOfBoundsException("extract_minor(int row, int col) error.
row > rowsize");
        };
        if (col > numofcols)
        {
            throw new IndexOutOfBoundsException("extract_minor(int row, int col) error.
col > colsize");
        };
        if (row < 0)
        {
            throw new IndexOutOfBoundsException("extract_minor(int row, int col) error.
row < 0");
        };
        if (col < 0)
        {
```

```
    throw new IndexOutOfBoundsException("extract_minor(int row, int col) error.
col < 0");
  };
  // for (0,0) do from (1 to row, 1 to col)
  if (row == 0 && col == 0)
  {
    for (int i = 1; i < numofrows; i = i + 1)
    {
      for (int j = 1; j < numofcols; j = j + 1)
      {
        minor.mArray[i - 1][j - 1] = mArray[i][j];
      };
    };
  };
  // for (row,0) do (0 to row - 1, 1 to col)
  if (row == numofrows && col == 0)
  {
    for (int i = 0; i < numofrows - 1; i = i + 1)
    {
      for (int j = 1; j < numofcols; j = j + 1)
      {
        minor.mArray[i][j - 1] = mArray[i][j];
      };
    };
  };
  // for (0, col) do (1 to row, 0 to col - 1)
  if (row == 0 && col == numofcols)
  {
    for (int i = 1; i < numofrows; i = i + 1)
    {
      for (int j = 0; j < numofcols - 1; j = j + 1)
      {
        minor.mArray[i - 1][j] = mArray[i][j];
      };
    };
  };
  // for (row, col) do (0 to row - 1, 0 to col - 1)
  if (row == numofrows)
  {
    if (col == numofcols)
    {
      for (int i = 0; i < numofrows - 1; i = i + 1)
      {
        for (int j = 0; j < numofcols - 1; j = j + 1)
        {
          minor.mArray[i][j] = mArray[i][j];
```

```
                };
            };
        };
    };
    // for (m, n)
    if (row < numofrows && col < numofcols)
    {
        // (0 to m - 1, 0 to n - 1)
        for (int i = 0; i < row - 1; i = i + 1)
        {
            for (int j = 0; j < col - 1; j = j + 1)
            {
                minor.mArray[i][j] = mArray[i][j];
            };
        };
        // (0 to m - 1, n + 1 to col)
        for (int i = 0; i < row - 1; i = i + 1)
        {
            for (int j = col + 1; j < numofcols; j = j + 1)
            {
                minor.mArray[i][j - 1] = mArray[i][j];
            };
        };
        // (m + 1 to row, 0 to n - 1)
        for (int i = row + 1; i < numofrows; i = i + 1)
        {
            for (int j = 0; j < col - 1; j = j + 1)
            {
                minor.mArray[i - 1][j] = mArray[i][j];
            };
        };
        // (m + 1 to row, n + 1 to col)
        for (int i = row + 1; i < numofrows; i = i + 1)
        {
            for (int j = col + 1; j < numofcols; j = j + 1)
            {
                minor.mArray[i - 1][j - 1] = mArray[i][j];
            };
        };
    };
    return minor;
}

public matrix[][] extract_minor()
/**
```

```
    description: complete extraction of minors in a matrix. e.g. if parameter is a 4x4
matrix, it will result in 4x4 matrix of minors of 3x3 each
    parameter: none
    return: M x N x X x Y 4-dimensional array. M x N will give extraction position of the
minor [ a(M, N) ], X x Y will give the minor matrix.
    algorithm: create a 2-dimensional array (M x N) of same size as this matrix. each of
the element of M x N will be handles to a minor of its
     *position. create each a 2-dimensional array to each of the element and evaluate the
minor for each
    */
    {
        matrix[][] CompleteMinor = new matrix[numofrows][numofcols];
        for (int i = 0; i < numofrows; i = i + 1)
        {
            for (int j = 0; j < numofcols; j = j + 1)
            {
                CompleteMinor[i][j] = new matrix(numofrows - 1, numofcols - 1);
                CompleteMinor[i][j] = extract_minor(i, j);
            };
        };
        return CompleteMinor;
    }

    public matrix cofactor() throws Exception
    {
        if (numofrows != numofcols)
        {
            throw new Exception("numofrows not equal numofcols. public matrix cofactor()
error");
        };
        float[][] f = new float[numofrows][numofcols];
        matrix[][] CMinor = new matrix[numofrows][numofcols];
        CMinor = extract_minor();
        for (int i = 0; i < numofrows; ++i)
        {
            for (int j = 0; j < numofcols; ++j)
            {
                f[i][j] = (float)(determinant(CMinor[i][j].mArray, numofrows - 1) *
Math.pow(-1.0, i + j));
            };
        };
        matrix cof_m = new matrix(f, numofrows, numofcols);
        return cof_m;
    }

    public matrix adjoint() throws Exception
```

```
    {
        if (numofrows != numofcols)
        {
            throw new Exception("numofrows not equal numofcols. public matrix adjoint()
error");
        };
        matrix m = new matrix(numofrows, numofcols);
        m = cofactor();
        // do transposition. from transpose()
        matrix temp = new matrix(numofcols, numofrows);
        for (int i = 0; i < numofrows; i = i + 1)
        {
            for (int j = 0; j < numofcols; j = j + 1)
            {
                temp.mArray[j][i] = m.mArray[i][j];
            };
        };
        return temp;
    }

    public matrix inverse() throws Exception
    {
        if (numofrows != numofcols)
        {
            throw new Exception("numofrows not equal numofcols. public matrix inverse()
error");
        };
        matrix m = new matrix(numofrows, numofcols);
        double det = determinant(mArray, numofrows);
        m = adjoint();
        for (int i = 0; i < numofrows; ++i)
        {
            for (int j = 0; j < numofcols; ++j)
            {
                m.mArray[i][j] = (float)(m.mArray[i][j] / det);
            };
        };
        return m;
    }

    public double determinant(float[][] m, int n) throws Exception
    {
        if (numofrows != numofcols)
        {
            throw new Exception("numofrows not equal numofcols. public double
determinant() error");
```

```
        };
        double det = 0.00;
        if (n == 1)
        {
            return m[0][0];
        };
        if (n == 2)
        {
            return (m[0][0] * m[1][1]) - (m[1][0] * m[0][1]);
        }
        else
        {
            //matrix[] m = new matrix[n];
            float[][][] f = new float[n][n - 1][n - 1];
            for (int i = 0; i < n; ++i)
            {
                //f = extract_minor(0, n);
                //m[i] = matrix(extract_minor(0,n).mArray, extract_minor(0,n).rowsize(),
extract_minor(0,n).colsize());
                f[i] = extract_minor(0, n).showMatrix();
                det = det + (Math.pow(-1.0, 1.0 + 1.0 + i) * m[0][i] * determinant(f[i], n - 1));
            };
        };
        return det;
    };
    // END OF OPERATION METHODS.
}
```

```
/*
 * matrix_operations.java
 *
 * This file contains the class matrix_operations.
 * Class matrix_operations holds all methods requiring more than one matrix.
 *
 * Created on February 1, 2003, 12:26 PM
 */

/**
 *
 * @author  Maurice@Server
 */

import java.lang.ArithmeticException;
```

```
public class matrix_operations {

    /** Creates a new instance of matrix_operations */
    public matrix_operations() {
    }
    public matrix addition(matrix p, matrix q) throws ArithmeticException
    /**
    description: method for matrix addition
    parameters: 2 2-dimensional matrices
    return: 1 2-dimensional matrix containing the sum
    algorithm: p = [ a(i,j) ]
            q = [ b(i,j) ]
            use 2 for loops to loop for row and column for addition
    */
    {
        matrix m = new matrix(p.rowsize(), p.colsize());
        if (p.rowsize() != q.rowsize())
        {
            throw new ArithmeticException("addition(matrix p, matrix q) error. row size
difference");
        };
        if (p.colsize() != q.colsize())
        {
            throw new ArithmeticException("addition(matrix p, matrix q) error. column size
difference");
        };
        for (int i = 0; i < m.rowsize(); i = i + 1)
        {
            for (int j = 0; j < m.colsize(); j = j + 1)
            {
                m.mArray[i][j] = p.mArray[i][j] + q.mArray[i][j];
            };
        };
        return m;
    }

    public matrix subtraction(matrix p, matrix q) throws ArithmeticException
    /**
    description: method for matrix subtraction
    parameters: 2 2-dimensional matrices
    return: 1 2-dimensional matrix containing the difference
    algorithm: p = [ a(i,j) ]
            q = [ b(i,j) ]
            use 2 for loops to loop for row and column for subtraction
    */
    {
```

```
        matrix m = new matrix(p.rowsize(), p.colsize());
        if (p.rowsize() != q.rowsize())
        {
            throw new ArithmeticException("subtraction(matrix p, matrix q) error. row size
difference");
        };
        if (p.colsize() != q.colsize())
        {
            throw new ArithmeticException("subtraction(matrix p, matrix q) error. column
size difference");
        };
        for (int i = 0; i < m.rowsize(); i = i + 1)
        {
            for (int j = 0; j < m.colsize(); j = j + 1)
            {
                m.mArray[i][j] = p.mArray[i][j] - q.mArray[i][j];
            };
        };
        return m;
    }

    public matrix multiplication(matrix p, matrix q) throws ArithmeticException
    /**
    description: method for matrix multiplication
    parameters: 2 2-dimensional matrices
    return: 1 2-dimensional matrix containing the product
    algorithm: transpose p
     *        do column multiplication using 2 for loops
    */
    {
        matrix m = new matrix(p.rowsize(), q.colsize());
        float sum = 0;
        p = p.transpose();
        for (int i = 0; i < p.colsize(); i = i + 1)
        {
            for (int j = 0 ; j < q.colsize(); j = j + 1)
            {
                for (int x = 0; x < p.rowsize(); x = x + 1)
                {
                    sum = sum + (p.mArray[x][i] * q.mArray[x][j]);
                };
                m.mArray[i][j] = sum;
            };
        };
        return m;
    }
```

```
    public matrix power(matrix p, int n)
    /**
     *description: method for evaluating power of matrix
     *parameters: a matrix and the n exponent
     *return: matrix of exponent
     *algorithm: similar to factorial evaluation but uses matrix multiplication method
     */
    {
        if (n == 1)
        {
            return p;
        }
        else
        {
            if (n == 2)
            {
                return multiplication(p, p);
            }
            else
            {
                return multiplication(p, power(p, n - 1));
            }
        }
    }


}
```

---

```
/*
 * StrToMatrix.java
 *
 * This file contains a parser class, StrToMatrix, which converts an input string to Matrix
class
 *
 * Created on January 31, 2003, 9:20 PM
 */

/**
 *
 * @author  Maurice@Server
 */

import java.lang.StringBuffer;
```

```
import java.util.StringTokenizer;

public class StrToMatrix {

    public String inStr;
    public StringBuffer inStrBuff;
    public StringTokenizer inStrTok;
    StringBuffer[] parseArray, tempArray;
    StringBuffer[][] full;
    float[][] preMatrix;
    public matrix m;

    /** Creates a new instance of StrToMatrix */
    // remove all trailing and leading white spaces
    // matrix comes in as string, eg. [[2 3 4]-[2 3 4]-[2 3 4]]
    public StrToMatrix(String s) {
        inStr = s.trim();
        inStrBuff = new StringBuffer(inStr);
        //inStrTok = new StringTokenizer(s.toString());
    }

    // THIS SECTION CONTAINS ALL CHECKING METHODS
    /*public int equalBracket(String s)
    /** Check for the number of open and close square brackets. Returns 1 if equal
     * 0 if not equal.
     */
    {
    }
    // END OF CHECKING METHODS

    // THIS SECTION CONTAINS ALL PARSING METHODS
    public StringBuffer deBracket(StringBuffer s)
    // removes the leftmost and rightmost bracket of the string.
    // after this method, matrix is [2 3 4]-[2 3 4]-[2 3 4]
    {
        //int index = s.length();
        //index = s.indexOf("[");
        s = s.deleteCharAt(s.indexOf("["));
        //index = s.lastIndexOf("]");
        s = s.deleteCharAt(s.lastIndexOf("]"));
        return s;
    }

    public StringBuffer[] parser(StringBuffer s)
    // this method breaks [2 3 4]-[2 3 4]-[2 3 4] into an array of
    // elements [2 3 4] and [2 3 4] and [2 3 4]
```

```
// there will be 1 more row than the number of "-"
{
   StringTokenizer inStrTok = new StringTokenizer(s.toString(), "-");
   int tokenize = inStrTok.countTokens();
   int count = 0;
   StringBuffer[] parseArray = new StringBuffer[tokenize];
   /*for (int i = 0; i < s.length(); i = i + 1)
   {
      if (s.charAt(i) == "-")
      {
         count = count + 1;
      };
   };
   StringBuffer[] parseArray = new StringBuffer[count + 1];*/
   while (inStrTok.hasMoreTokens())
   {
      parseArray[count] = new StringBuffer(inStrTok.nextToken());
      count = count + 1;
   };
   return parseArray;
}

public StringBuffer[][] arrayParse()
{
   StringBuffer[] tempArray = new StringBuffer[parseArray.length];
   StringTokenizer t = new StringTokenizer(parseArray[1].toString());
   StringBuffer[][] full = new StringBuffer[parseArray.length][t.countTokens()];
   StringBuffer[] temp = new StringBuffer[t.countTokens()];
   tempArray = parseArray;
   for (int i = 0; i < tempArray.length; i = i + 1)
   {
       full[i] = parser(tempArray[i]);
   };
   return full;
}

public float[][] StrBuffArrToFloatArr()
// converts all the entities of full array (StringBuffer) into a float array
{
   //Float tempf = new Float(0.00);
   for (int i = 0; i < full.length; i = i + 1)
   {
      for (int j = 0; j < full[1].length; j = j + 1)
      {
         //String y = full[i][j].toString();
         preMatrix[i][j] = Float.parseFloat(full[i][j].toString());
```

```
                //preMatrix[i][j] = tempf;
            };
        };
        return preMatrix;
    }

    public matrix writeMatrix()
    {
        inStrBuff = deBracket(inStrBuff);
        parseArray = parser(inStrBuff);
        for (int i = 0; i < parseArray.length; i = i + 1)
        {
            parseArray[i] = deBracket(parseArray[i]);
        };
        full = arrayParse();
        preMatrix = StrBuffArrToFloatArr();
        m = new matrix(preMatrix, preMatrix.length, preMatrix[1].length);
        return m;
    }

/*  public matrix writeMatrix()
    {
        inStrBuff = deBracket(inStrBuff);
        tempArray = parser(s);
        float[][] x = full;
        for (int i = 0; i < x.length; i = i + 1)
        {
            for (int j = 0; j < x[1].length; j = j + 1)
            {
                m.mArray[i][j] =
    }
/*  public matrix writeMatrix()
    {
        StringBuffer[][] m = new StringBuffer[inStrTok.countTokens()]
                              [inStrTok.nextToken().countTokens()];
        float[][] f = new float[inStrTok.countTokens()]
                              [inStrTok.nextToken().countTokens()];
        int count;
        inStrBuff = deBracket(inStrBuff);
        parseArray = parser(inStrBuff);
        for (int i = 0; i < parseArray.length; i = i + 1)
        {
            m[i][0] = deBracket(parseArray[i]);
            while (inStrTok.nextToken().hasMoreTokens())
            {
                m[i][count] = new StringBuffer(inStrTok.nextToken().nextToken());
```

```
                count = count + 1;
            };
        };
        for (int j = 0; j < f.length; j = j + 1)
        {
            for (int k = 0; k < f[1].length; k = k + 1)
            {
                f[j][k] = float.valueOf(m[j][k].toString());
            };
        };
        return matrix(f, j, k);
    }
 */
    // END OF PARSING METHODS
}
```

---

```
/*
 * MatrixToStr.java
 *
 * This file contains an assembler class, MatrixToStr, which converts a matrix back to a
string
 *
 * Created on February 12, 2003, 11:56 PM
 */

/**
 *
 * @author  Maurice@Server
 */

import java.lang.StringBuffer;

public class MatrixToStr {

    public float[][] inFloat;
    public StringBuffer[][] iniStrBuff;
    public StringBuffer[] assStrBuff;
    public StringBuffer fullStrBuff;

    /** Creates a new instance of MatrixToStr */
    public MatrixToStr(matrix m)
    {
        inFloat = m.mArray;
    }
```

```
public StringBuffer[][] FloatArrToStrBuffArr()
// converts all elements of full float array to StringBuffer array
{
   StringBuffer[][] iniStrBuff = new StringBuffer[inFloat.length][inFloat[1].length];
   for (int i = 0; i < iniStrBuff.length; i = i + 1)
   {
      for (int j = 0; j < iniStrBuff[1].length; ++j)
      {
         iniStrBuff[i][j] = new StringBuffer(Float.toString(inFloat[i][j]));
      };
   };
   return iniStrBuff;
}

public StringBuffer[] arrayAssemble()
// assembles a 2-d stringbuffer array into a 1-d stringbuffer array with " " as delimiters
{
   StringBuffer[] assStrBuff = new StringBuffer[iniStrBuff.length];
   for (int i = 0; i < iniStrBuff.length; i = i + 1)
   {
      for (int j = 0; j < iniStrBuff[1].length; ++j)
      {
         if (j < iniStrBuff[1].length)
         {
            assStrBuff[i] = assStrBuff[i].append(iniStrBuff[i][j]);
         }
         else
         {
            assStrBuff[i] = assStrBuff[i].append(iniStrBuff[i][j]);
            assStrBuff[i] = assStrBuff[i].append(" ");
         };
      };
   };
   return assStrBuff;
}

public StringBuffer reBracket(StringBuffer s)
// flank stringbuffer with "[" and "]"
{
   s = s.append("]");
   s = s.insert(0, "[");
   return s;
}

public StringBuffer fullAssemble()
```

```
    {
        StringBuffer fullStrBuff = new StringBuffer();
        for (int i = 0; i < assStrBuff.length; i = i + 1)
        {
            if (i < assStrBuff.length)
                {
                    fullStrBuff = fullStrBuff.append(assStrBuff[i]);
                }
                else
                {
                    fullStrBuff = fullStrBuff.append(assStrBuff[i]);
                    fullStrBuff = fullStrBuff.append("-");
                };
        }
        return fullStrBuff;
    }

    public String writeString()
    {
        iniStrBuff = FloatArrToStrBuffArr();
        assStrBuff = arrayAssemble();
        for (int i = 0; i < assStrBuff.length; ++i)
        {
            assStrBuff[i] = reBracket(assStrBuff[i]);
        };
        fullStrBuff = fullAssemble();
        fullStrBuff = reBracket(fullStrBuff);
        return fullStrBuff.toString();
    }
}
```

| End of Source Codes |
|---|

# APPENDIX B:
# PACKAGE DOCUMENTATION

Package Class **Tree** **Deprecated** **Index** **Help**

PREV NEXT FRAMES NO FRAMES All Classes

# Hierarchy For All Packages

## Class Hierarchy

- class java.lang.Object
  - class **matrix**
  - class **matrix_operations**
  - class **MatrixToStr**
  - class **StrToMatrix**

Package **Class** **Tree** **Deprecated** **Index** **Help**
PREV CLASS **NEXT CLASS** **FRAMES** **NO FRAMES** **All Classes**
SUMMARY: NESTED | FIELD | CONSTR | METHOD DETAIL: FIELD | CONSTR | METHOD

# Class matrix

```
java.lang.Object
  |
  +-matrix
```

public class **matrix**
extends java.lang.Object

| Constructor Summary |
|---|
| **matrix**(float[][] m, int row, int col)<br><br>Creates a matrix of m. |
| **matrix**(int row, int col)<br><br>Creates an empty matrix of number of rows and columns as specified. |

| Method Summary | |
|---|---|
| matrix | **adjoint**()<br><br>Performs adjoint operation. |
| matrix | **cofactor**()<br><br>Performs cofactoring operation. |
| int | **colsize**()<br><br>Returns the column size of the matrix. |
| double | **determinant**(float[][] m, int n)<br><br>Returns the determinant of m matrix of size n. |
| matrix[][] | **extract_minor**()<br><br>Performs minor operation on every element in the matrix. A 4 x 4 matrix will yield 16 minors. |
| matrix | **extract_minor**(int row, int col)<br><br>Performs minor operation on the given element. |
| void | **fillmatrix**(float[][] m)<br><br>Method to replace or to assign a matrix. |
| matrix | **inverse**()<br><br>Performs inverse operation on the matrix. |
| int | **isIdentity**()<br><br>Checks if the matrix is an identity matrix. |
| int | **isSquare**()<br><br>Checks if the matrix is a square matrix. |
| int | **rowsize**()<br><br>Returns the number of rows on the matrix. |
| float[][] | **showMatrix**()<br><br>Dereference the matrix. |
| matrix | **transpose**()<br><br>Performs transposition operation on the matrix. |

| **Methods inherited from class java.lang.Object** |
|---|
| clone, equals, finalize, getClass, hashCode, notify, notifyAll, toString, wait, wait, wait |

| **Constructor Detail** |
|---|

### matrix

Creates an empty matrix of row size and column size

```
public matrix(int row,
              int col)
       throws java.lang.NegativeArraySizeException
```

NegativeArraySizeException is thrown when either the row or col is not a positive integer.

### matrix

Creates a matrix of m

```
public matrix(float[][] m,
              int row,
              int col)
       throws java.lang.ArrayIndexOutOfBoundsException,
              java.lang.NegativeArraySizeException
```

ArrayIndexOutOfBoundsException is thrown when either the row or col size do not match that of array m.
NegativeArraySizeException is thrown when either the row or col is not a positive integer.

## Method Detail

### fillmatrix

```
public void fillmatrix(float[][] m)
                throws java.lang.ArrayIndexOutOfBoundsException
```

java.lang.ArrayIndexOutOfBoundsException

ArrayIndexOutOfBoundsException is thrown when either the row or col size (that is previously defined) do not match that of array m.

### rowsize

Returns the row size of the matrix.

```
public int rowsize()
```

### colsize

Returns the column size of the matrix.

```
public int colsize()
```

### showMatrix

Dereference the matrix.

```
public float[][] showMatrix()
```

### isSquare

Checks if the matrix is a square matrix.

```
public int isSquare()
```

---

### isIdentity

Checks if the matrix is an identity matrix.

```
public int isIdentity()
```

---

### transpose

Performs transposition operation of the matrix.

```
public matrix transpose()
```

---

### extract_minor

Performs minor operation on the given element.

```
public matrix extract_minor(int row,
                            int col)
                     throws java.lang.IndexOutOfBoundsException
      java.lang.IndexOutOfBoundsException

      IndexOutOfBoundsException is thrown when either the row or col do
      not falls within the limits of the matrix.
```

---

### extract_minor

Performs minor operation on every element in the matrix. A 4 x 4 matrix will yield 16 minors.

```
public matrix[][] extract_minor()
```

---

### cofactor

Performs cofactoring operation.

```
public matrix cofactor()
                throws java.lang.Exception
      java.lang.Exception

      Exception is thrown when attempt to perform this operation on a
      non-square matrix.
```

---

## adjoint

Performs adjoint operation.

```
public matrix adjoint()
               throws java.lang.Exception
```

java.lang.Exception

Exception is thrown when attempt to perform this operation on a non-square matrix.

---

## inverse

Performs inverse operation on the matrix.

```
public matrix inverse()
               throws java.lang.Exception
```

java.lang.Exception

Exception is thrown when attempt to perform this operation on a non-square matrix.

---

## determinant

Returns the determinant of m matrix of size n.

```
public double determinant(float[][] m,
                          int n)
                   throws java.lang.Exception
```

java.lang.Exception

Exception is thrown when attempt to perform this operation on a non-square matrix.

---

Package **Class** **Tree** **Deprecated** **Index** **Help**

PREV CLASS **NEXT CLASS** **FRAMES** **NO FRAMES** **All Classes**

SUMMARY: NESTED | FIELD | CONSTR | METHOD DETAIL: FIELD | CONSTR | METHOD

Package **Class** **Tree** **Deprecated** **Index** **Help**

PREV CLASS NEXT CLASS **FRAMES** **NO FRAMES** **All Classes**

SUMMARY: NESTED | FIELD | CONSTR | METHOD DETAIL: FIELD | CONSTR | METHOD

# Class matrix_operations

java.lang.Object
  |
  +-**matrix_operations**

public class **matrix_operations**
extends java.lang.Object

| Constructor Summary |
|---|
| **matrix_operations**()<br>Creates a new instance of matrix_operations. |

| Method Summary | |
|---|---|
| `matrix` | **addition**(`matrix p, matrix q`)<br><br>Adds P and Q and exports the sum. |
| `matrix` | **multiplication**(`matrix p, matrix q`)<br><br>Multiplies P to Q (P.Q) and exports the product. |
| `matrix` | **power**(`matrix p, int n`)<br><br>Multiplies P to the power of n. |
| `matrix` | **subtraction**(`matrix p, matrix q`)<br><br>Subtracts P and Q (P – Q) and exports the difference. |

| Methods inherited from class java.lang.Object |
|---|
| `clone, equals, finalize, getClass, hashCode, notify, notifyAll,` |

| toString, wait, wait, wait |
|---|

## Constructor Detail

### matrix_operations

public **matrix_operations**()

Creates a new instance of matrix_operations

## Method Detail

### addition

Adds `P and Q and exports the sum.`

public matrix **addition**(matrix p,
matrix q)
throws java.lang.ArithmeticException

```
java.lang.ArithmeticException

ArithmeticException is thrown when either the row size of the 2
matrices are different (Error Message: addition(matrix p, matrix
q) error. row size difference) or the column size of the 2
matrices are different (Error Message: addition(matrix p,
matrix q) error. column size difference).
```

---

### subtraction

```
Subtracts P and Q (P – Q) and exports the difference.
```

public matrix **subtraction**(matrix p,
matrix q)
throws java.lang.ArithmeticException

```
java.lang.ArithmeticException

ArithmeticException is thrown when either the row size of the 2
matrices are different (Error Message: subtraction(matrix p,
matrix q) error. row size difference) or the column size of
the 2 matrices are different (Error Message:
subtraction(matrix p, matrix q) error. column size
difference).
```

---

### multiplication

```
Multiplies P to Q (P.Q) and exports the product.
```

public matrix **multiplication**(matrix p,

matrix q)
throws java.lang.ArithmeticException

`java.lang.ArithmeticException`

ArithmeticException is thrown when the column size of P is different from the row size of Q (Error Message: multiplication(matrix p, matrix q) error.).

---

### power

`Multiplies P to the power of n.`

public matrix **power**(matrix p,
int n)

---

Package **Class** **Tree** **Deprecated** **Index** **Help**

PREV CLASS NEXT CLASS **FRAMES** **NO FRAMES** **All Classes**

SUMMARY: NESTED | FIELD | CONSTR | METHOD DETAIL: FIELD | CONSTR | METHOD

---

# Class MatrixToStr

```
java.lang.Object
  |
  +-MatrixToStr
```

public class **MatrixToStr**
extends java.lang.Object

| **Field Summary** | |
|---|---|
| java.lang.StringBuffer[] | **assStrBuff**<br><br>Row assembly array of iniStrBuff. |
| java.lang.StringBuffer | **fullStrBuff**<br><br>Full assembly from assStrBuff. |
| float[][] | **inFloat**<br><br>Input array of float. |
| java.lang.StringBuffer[][] | **iniStrBuff**<br><br>Converted array of float (to StringBuffer) |

| **Constructor Summary** |
|---|
| **MatrixToStr**(matrix m)<br>Creates a new instance of MatrixToStr |

| **Method Summary** | |
|---|---|
| java.lang.StringBuffer[] | **arrayAssemble**()<br><br>Takes from iniStrBuff, and performs row assembly by merging the column, with space as delimiter. |
| java.lang.StringBuffer[][] | **FloatArrToStrBuffArr**()<br><br>Method to convert a two-dimensional float array into a two-dimensional StringBuffer array. |
| java.lang.StringBuffer | **fullAssemble**()<br><br>Takes from assStrBuff, and performs assembly by concatenation of assStrBuff[i], where i = 0 to n, with "-" as delimiter. |
| java.lang.StringBuffer | **reBracket**(java.lang.StringBuffer s)<br><br>Method to flank s with "[" and "]". |
| java.lang.String | **writeString**()<br><br>Method to call for writing a two-dimensional float array to a String type after constructor method. This method will chain all methods as needed. |

| **Methods inherited from class java.lang.Object** |
|---|
| clone, equals, finalize, getClass, hashCode, notify, notifyAll, toString, wait, wait, wait |

## Field Detail

### inFloat

Input array of float.

`public float[][] inFloat`

---

### iniStrBuff

Converted array of float (to StringBuffer)

`public java.lang.StringBuffer[][] iniStrBuff`

---

### assStrBuff

Row assembly array of iniStrBuff.

```
public java.lang.StringBuffer[] assStrBuff
```

---

### fullStrBuff

```
Full assembly from assStrBuff.
```

```
public java.lang.StringBuffer fullStrBuff
```

## Constructor Detail

### MatrixToStr

```
public MatrixToStr(matrix m)
```

Creates a new instance of MatrixToStr

## Method Detail

### FloatArrToStrBuffArr

```
Method to convert a two-dimensional float array into a two-dimensional
StringBuffer array.
```

```
public java.lang.StringBuffer[][] FloatArrToStrBuffArr()
```

---

### arrayAssemble

```
Takes from iniStrBuff, and performs row assembly by merging the column,
with space as delimiter.
```

```
public java.lang.StringBuffer[] arrayAssemble()
```

---

### reBracket

```
Method to flank s with “[“ and “]”.
```

```
public java.lang.StringBuffer reBracket(java.lang.StringBuffer s)
```

---

### fullAssemble

```
Takes from assStrBuff, and performs assembly by concatenation of
assStrBuff[i], where i = 0 to n, with “-“ as delimiter.
```

```
public java.lang.StringBuffer fullAssemble()
```

---

### writeString

```
Method to call for writing a two-dimensional float array to a String
type after constructor method. This method will chain all methods as
needed.

public java.lang.String writeString()
```

Package **Class** **Tree** **Deprecated** **Index** **Help**

**PREV CLASS** **NEXT CLASS** **FRAMES** **NO FRAMES** **All Classes**

SUMMARY: NESTED | FIELD | CONSTR | METHOD DETAIL: FIELD | CONSTR | METHOD

Package **Class** **Tree** **Deprecated** **Index** **Help**

**PREV CLASS** NEXT CLASS
SUMMARY: NESTED | FIELD | CONSTR | METHOD

**FRAMES** **NO FRAMES** **All Classes**
DETAIL: FIELD | CONSTR | METHOD

# Class StrToMatrix

```
java.lang.Object
  |
  +-StrToMatrix
```

public class **StrToMatrix**
extends java.lang.Object

## Field Summary

| | |
|---|---|
| java.lang.String | **inStr**<br><br>Input String type. |
| java.lang.StringBuffer | **inStrBuff**<br><br>StringBuffer representation of inStr. |
| java.util.StringTokenizer | **inStrTok**<br><br>Converted from inStrBuff for tokenizing. |
| matrix | **m**<br><br>Assembled matrix. |

## Constructor Summary

| |
|---|
| **StrToMatrix**(java.lang.String s)<br>Creates a new instance of StrToMatrix |

## Method Summary

| | |
|---|---|
| java.lang.StringBuffer[][] | **arrayParse**()<br><br>Method to tokenize into a two-dimensional array of StringBuffer. |
| java.lang.StringBuffer | **deBracket**(java.lang.StringBuffer s)<br><br>Method to remove flanking "[" and "]". |
| java.lang.StringBuffer[] | **parser**(java.lang.StringBuffer s)<br><br>Method to tokenize s into an array of StringBuffer by "-" as delimiter. |
| float[][] | **StrBuffArrToFloatArr**()<br><br>Method to convert a two-dimensional StringBuffer array into a two-dimensional float array. |
| matrix | **writeMatrix**()<br><br>Method to call for writing a String into a matrix after constructor method. This method will chain all methods as needed. |

| **Methods inherited from class java.lang.Object** |
|---|
| clone, equals, finalize, getClass, hashCode, notify, notifyAll, toString, wait, wait, wait |

| **Field Detail** |
|---|

### inStr

Input String type.

public java.lang.String **inStr**

---

### inStrBuff

StringBuffer representation of inStr.

public java.lang.StringBuffer **inStrBuff**

---

### inStrTok

Converted from inStrBuff for tokenizing.

`public java.util.StringTokenizer` **`inStrTok`**

---

### m

`Assembled matrix.`

`public matrix` **`m`**

## Constructor Detail

### StrToMatrix

`public` **`StrToMatrix`**`(java.lang.String s)`

Creates a new instance of StrToMatrix

## Method Detail

### deBracket

`Method to remove flanking “[“ and “]”.`

`public java.lang.StringBuffer` **`deBracket`**`(java.lang.StringBuffer s)`

---

### parser

`Method to tokenize s into an array of StringBuffer by “-“ as delimiter.`

`public java.lang.StringBuffer[]` **`parser`**`(java.lang.StringBuffer s)`

---

### arrayParse

`Method to tokenize into a two-dimensional array of StringBuffer.`

`public java.lang.StringBuffer[][]` **`arrayParse`**`()`

---

### StrBuffArrToFloatArr

Method to convert a two-dimensional StringBuffer array into a two-dimensional float array.

`public float[][]` **`StrBuffArrToFloatArr`**`()`

---

### writeMatrix

Method to call for writing a String into a matrix after constructor method. This method will chain all methods as needed.

```
public matrix writeMatrix()
```

Package **Class** Tree Deprecated Index Help

PREV CLASS NEXT CLASS

FRAMES NO FRAMES All Classes

SUMMARY: NESTED | FIELD | CONSTR | METHOD

DETAIL: FIELD | CONSTR | METHOD

# A

# C

# D

# E

## F

---

## I

---

## M

---

# P

---

# R

---

# S

---

# T

---

# W

A C D E F I M P R S T W

Package Class **Tree** **Deprecated** **Index** **Help**

PREV NEXT **FRAMES** **NO FRAMES** **All Classes**